\documentclass[12pt]{article}
\newcommand{\be}{\begin{equation}}
\newcommand{\ee}{\end{equation}}
\newcommand{\bea}{\begin{eqnarray}}
\newcommand{\eea}{\end{eqnarray}}
\begin{document}
\title{Lower bounds of success probabilities for high-fidelity approach in KLM scheme
  }
\author{Kazuto Oshima\thanks{E-mail: oshima@nat.gunma-ct.ac.jp}    \\ \\
\sl National  Institute of Technology,  Gunma College,  Maebashi 371-8530, Japan }

\date{}
\maketitle
\begin{abstract}
In the Knill-Laflamme-Milburn (KLM) scheme, the success probability of quantum teleportation is given by
${n \over n+1}$, wehre $2n$ is the number of the ancilla qubits. For the high-fidelity approach in the KLM
scheme, the success probability is approximately given by $1-{1 \over n^{2}}$ for large $n$.  We give
an explicit prescription to prepare an optimal ancilla state and give an exact lower bound of the success probability for the high-fidelity approach for arbitrary $n$. 
\end{abstract}

PACS number(s): 03.67.Lx\\
\newpage
\section{Introduction}
Quantum teleportation\cite{Bennett} is the most important technique in quantum computation
and quantum communication.  Cluster state quantum computation\cite{Briegel} is a useful application
of quantum teleportation.  Photon is expected be one of the most promising media for quantum information.
Using  photons as qubits is advantageous in the points that it is robust against noise and it is easily 
transmitted.  It is, however, difficult to perform two-qubits gates between two photonic qubits.
 Knill, Laflamme and Milburn(KLM) \cite{Knill} have invented a scheme to carry out quantum teleportation
of a photonic qubit with the probability near to 1.  Combining the KLM scheme and the strategy of
Gottesman and Chuang\cite{Gottesman}, we can perform two-qubits gates between two photonic
qubits with the probability near to 1.   Nearly deterministic  two-qubits gates are useful in connecting cluster states
into arbitrary shapes as well as in universal quantum computation.
In the KLM scheme, to boost the success probability close to 1, we should prepare a large number of ancilla qubits.  This causes the difficulty that that the number optical resources such as
beam splitters and phase shifters increases.  Optimality about ancilla states have been studied by  Grudka and  Modlawska\cite{Grudka} for the purpose of perfect quantum teleportation.

Franson et. al. \cite{Franson} have proposed a new line of approach (high-fidelity approach) in the KLM scheme.   Preparing ancilla states properly, they have shown that the success
probability, in the sense of fidelity, approaches to 1 faster than the original setting  as the number of ancilla qubits increases.   Their approach is useful in saving the resources.  Their analysis, however, uses approximation
valid when the number of ancilla qubits is sufficiently large.   The purpose of this paper is to give an 
explicit prescription for preparing an optimal ancilla state for any number of ancilla qubits.  This ancilla
state gives a lower bound of the success probability of quantum teleportation without approximation in the sense of fidelity.

\section{Ancilla states and success probabilities} 
We prepare a $2n$-qubits ancilla state as $|u_{n}\rangle=\Sigma_{i=0}^{n}f(i)|0\rangle^{n-i}|1\rangle^{i}|0\rangle^{i}
|1\rangle^{n-i}$, where $|0\rangle^{i}$ means $i$ photons in the state $|0\rangle$ etc. and  $f(i)$'s are real
coefficients normalized as $\Sigma_{i=0}^{n}f(i)^{2}=1$.  In the original KLM scheme they are settled as $f(i)={1 \over \sqrt{n+1}},i=0,1,\cdots n$. We consider to teleport a quantum state $|\psi\rangle=\alpha|0\rangle +\beta|1\rangle, |\alpha|^{2}+|\beta|^{2}=1$. We perform $n+1$-poin quantum Fourier transformation ${\hat F}_{n+1}$ on the state $|\psi\rangle$ and the first $n$ qubits in the ancilla state.  Suppose we observe $k(0 \le k \le n+1)$ photons
after the transformation.  When $k=0$ and $k=n+1$ we fail in the procedure.   In another case, we obtain the teleported state  
\begin{equation}
 |q_{k}\rangle={{\alpha}f(k)|0\rangle+{\beta}f(k-1)|1\rangle \over \sqrt{|\alpha|^{2}f(k)^{2}+|\beta|^{2}f(k-1)^{2}}}
\label{kstate}
\end{equation}
at the $k$-th qubit in the latter half of the ancilla qubits.   To obtain the state $|q_{k}\rangle$ in the form of Eq.(1), we should perform certain 
relative phase shift between the states $|0\rangle$ and $|1\rangle$ depending
on the observed $k$-photon state.  The probability $p_{k}$
to obtain the state $|q_{k}\rangle$ is given by
\begin{equation}
p_{k}=\Sigma_{k}| \langle k|{\hat F_{n+1}}|\psi\rangle|u_{n}\rangle|^{2}=|\alpha|^{2}f(k)^{2}+|\beta|^{2}f(k-1)^{2},
\label{kprobability}
\end{equation}
where the summation about $k$ runs over all possible $k$-photon states and we have used $\Sigma_{k}|k\rangle\langle{k}|={\hat I}_{k}$ with ${\hat I}_{k}$ 
the identity operator on any $k$-photon state.  In the
high-fidelity approach the success probability $p$ is defined by the expectation value of the square of the 
fidelity $|\langle \psi|q_{k}\rangle|^{2}, k=1,2,\cdots, n$; we define $p$ as
$p=\sum_{k=1}^{n}p_{k}|\langle \psi|q_{k}\rangle|^{2}$.

First, let us choose the original state $|\psi\rangle$ as $|\psi\rangle=|+\rangle={ 1\over \sqrt{2}}(|0\rangle+|1\rangle)$. 
 In this case the fidelity $F_{k}$ is given by
\begin{equation}
F_{k}= |\langle +|q_{k}\rangle|={1 \over \sqrt{2}}{f(k)+f(k-1) \over \sqrt{f(k)^{2}+f(k-1)^{2}}}
\label{kfidelity}
\end{equation}
and the success probability $p$ is given by
\begin{equation}
p={1 \over 4}\Sigma_{k=1}^{n}(f(k)+f(k-1))^{2}.
\label{pplus}
\end{equation}
The maximum value of $p$ is given by the largest eigenvalue $\lambda_{n}$ of the following $(n+1)\times(n+1)$
matrix
\begin{eqnarray}
A={1 \over 4}\left(\begin{array}{cccccc}
       1 & 1 & 0 & \ldots &  0 &0\\
       1 & 2 & 1 &  \ddots& 0 & 0\\
       0  & 1 & 2  &  \ddots &  0 & \vdots    \\
       \vdots  & \ldots & \ddots  & \ddots & 1&0 \\
       0 & \ldots & \ldots  & 1 & 2 & 1  \\
       0 &  \ldots &    \dots & 0 & 1 & 1    \end{array} \right). 
\label{matrix}
\end{eqnarray}
This matrix has recently appeared in the literature\cite{Sergii} in the context of port-based teleportation
scheme\cite{Hiroshima}. 
The corresponding normalized eigenvector of $A$ gives the optimal coefficients
\begin{equation}
(f(0),f(1), \cdots, f(n)). 
\label{coefficints}
\end{equation}
 Note that the condition $f(0)=f(n)$ holds, on account of the symmetric properties of the matrix $A$.

Second, we consider an arbitrary quantum state $|\psi\rangle=\alpha|0\rangle+\beta|1\rangle$.
Having settled the coefficients $f(k)$'s as above, the success probability for the state  $|\psi\rangle=\alpha|0\rangle+\beta|1\rangle$ is giben by
\begin{equation}
p=\Sigma_{k=1}^{n}(|\alpha|^{2}f(k)+|\beta|^{2}f(k-1))^{2}.
\label{probability}
\end{equation}
It is easy to see that the success probability $p$ takes its minimum value $\lambda_{n}$ at
$|\alpha|^{2}=|\beta|^{2}={1 \over 2}$ as far as the condition $f(0)=f(n)$ is satisfied. This means that
the success probability $p$ is bounded from below by  $\lambda_{n}$ for arbitrary quantum state
 $|\psi\rangle=\alpha|0\rangle+\beta|1\rangle$. For references, we exhibit the value $\lambda_{n}$
and the corresponding coefficients $f(k)$'s for some  small $n$'s; $\lambda_{2}={3 \over 4}$, $ (f(0),f(1),f(2))=
{1 \over \sqrt{6}}(1,2,1)$;  $\lambda_{3}={2+\sqrt{2} \over 4}$, $ (f(0),f(1),f(2),f(3))=
{1 \over \sqrt{8+4\sqrt{2}}}(1,1+\sqrt{2},1+\sqrt{2},1)$; $\lambda_{4}={5+\sqrt{5} \over 8}$, $ (f(0),f(1),f(2),f(3),f(4))=
{1 \over \sqrt{16+5\sqrt{5}}}(1,{3+\sqrt{5} \over 2},1+\sqrt{5},{3+\sqrt{5} \over 2} ,1)$. 
It will be not so hard to see that $\lambda_{n}$ approaches to 1 as  $\lambda_{n}={1 \over 2}+{1 \over 2}\cos{\pi \over n+1}$ as $n$ increases. 
 The largest eigenvalue $\lambda_{n}$ can also be written as $\lambda_{n}={1 \over 2}+{1 \over 4}\mu_{n}$, where $\mu_{n}=2\cos{\pi \over n+1}$ 
is the largest eigenvalue of the following simple $n \times n$ matrix $B$:  
\begin{eqnarray}
B= \left(\begin{array}{cccccc}
       0 & 1 & 0 & \ldots &  0 &0\\
       1 & 0 & 1 &  \ddots& \vdots & 0\\
       0  & 1 & 0  &  \ddots &  0 & \vdots    \\
       \vdots  & \ddots & \ddots  & \ddots & 1&0 \\
       0 & \ldots & \ddots  & 1 & 0 & 1  \\
       0 &  \ldots &    \dots & 0 & 1 & 0    \end{array} \right). 
\label{matrixB}
\end{eqnarray}
An expectation value of fidelity, not the square of fidelity, has already been computed in closed form
in the port-based teleportation scheme, which is somewhat different from the present one,
with an elaborate method\cite{Ishizaka}.
\\

\section{Conclusions}
We have given the exact lower bound of the success probability of quantum teleportation for high-fidelity
approach in the KLM scheme.  Our result is valid for any number of ancilla qubits and is of practical use.  We have given the 
prescription to prepare the most preferable anncilla state for any number of ancilla qubits.
\\
\\
{\bf Acknowledgment}\\
The author thanks Kenta Takahashi and Yuya Tamura for useful discussion. 
The author thanks Sergii Strelchuk for helpful advice.

\end{document}